\documentclass[11pt,a4paper]{report}
\usepackage[utf8]{inputenc}
\usepackage{multirow}
\usepackage{rotating}

\usepackage[top=1in, bottom=1.25in, left=1in, right=1in]{geometry}

\usepackage{microtype}
\usepackage{lineno}  % for line numbering during review
\usepackage{xspace} % To avoid problems with missing or double spaces after
\usepackage{caption} %these three command get the figure and table captions automatically small

\usepackage{graphicx}  % to include figures (can also use other packages)
\usepackage{color}
\usepackage{colortbl}
\graphicspath{{./figs/}} % Make Latex search fig subdir for figures
\DeclareGraphicsExtensions{.pdf,.PDF,png,.PNG}

\usepackage{amsmath} % Adds a large collection of math symbols
\usepackage{amssymb}
\usepackage{amsfonts}
\usepackage{upgreek} % Adds in support for greek letters in roman typeset

\newcommand*\patchAmsMathEnvironmentForLineno[1]{%
\expandafter\let\csname old#1\expandafter\endcsname\csname #1\endcsname
\expandafter\let\csname oldend#1\expandafter\endcsname\csname
end#1\endcsname
 \renewenvironment{#1}%
   {\linenomath\csname old#1\endcsname}%
   {\csname oldend#1\endcsname\endlinenomath}%
}
\newcommand*\patchBothAmsMathEnvironmentsForLineno[1]{%
  \patchAmsMathEnvironmentForLineno{#1}%
  \patchAmsMathEnvironmentForLineno{#1*}%
}
\AtBeginDocument{%
\patchBothAmsMathEnvironmentsForLineno{equation}%
\patchBothAmsMathEnvironmentsForLineno{align}%
\patchBothAmsMathEnvironmentsForLineno{flalign}%
\patchBothAmsMathEnvironmentsForLineno{alignat}%
\patchBothAmsMathEnvironmentsForLineno{gather}%
\patchBothAmsMathEnvironmentsForLineno{multline}%
\patchBothAmsMathEnvironmentsForLineno{eqnarray}%
}

\usepackage{hyperxmp}

\usepackage[pdftex,
            plainpages=false,
            pdfpagelabels]{hyperref}

\usepackage[colorinlistoftodos,textsize=scriptsize]{todonotes}

\usepackage[all]{hypcap} % Internal hyperlinks to floats.

\usepackage{cite} % Allows for ranges in citations
\usepackage{mciteplus}

\begin{document}
%\linenumbers

%%%%%%%%%%%%%%%%%%%%%%%%%
%%%%%  TITLE PAGE  %%%%%%
%%%%%%%%%%%%%%%%%%%%%%%%%
\begin{titlepage}
\vspace*{3cm}

% Title --------------------------------------------------
{\bf\boldmath\LARGE
\begin{center}
LPNHE scientific perspectives for the \\ European Strategy for Particle Physics
\end{center}
}

\vspace*{2cm}

% Authors -------------------------------------------------
%\begin{center}
%\Large
 This note summarizes the activities and the scientific and technical perspectives of the Laboratoire de Physique Nucleaire et de Hautes Energies (LPNHE) at Sorbonne University, Paris. Although the ESPP is specifically aimed at particle physics, we discuss in this note in parallel the three scientific lines developed at LPNHE (Particle Physics, Astroparticles, Cosmology), first with the current scientific activities, then for the future activities. However, our conclusions and recommendations are focused on the particle physics strategy. \\ 
 \bigskip
 \begin{center}
 %Everyone in the lab who has contributed to this document or wishes to be associated with its content is welcome to sign it. \\

  \bigskip
 %Please send a mail to Vladimir.Gligorov@cern.ch or any of the editorial team members (Reina Camacho, Nicolas Busca, Mathieu Guigue, Romain Ga\"{i}or, Marco Bomben) by thursday to add your name to the list.\\
 \bigskip
 E.~Ben~Haim, G.~Bernardi, E.~Bertholet, J.~Bolmont, M.~Bomben, N.~Busca, G.~Calderini, R.~Camacho~Toro, M.~Charles, J.~Chauveau, R.~Cornat, F.~Crescioli, J.~Da~Rocha, L.~D’Eramo, L.~Delbuono, F.~Derue, R.~Ga\"{i}or, C.~Giganti, V.~V.~Gligorov, M.~Guigue, F.~Kapusta, L.~Khalil, D.~Lacour, B.~Laforge, J-P.~Lenain, A.~Letessier-Selvon, K.~Liu, E.~Lopez~Fune, B.~Malaescu, O.~Martineau, G.~Marchiori, I.~Nikolic-Audit, I.~Nomidis, J.~Ocariz, L.~Pascual-Dominguez, F.~Polci, P.~Privitera, A.~Robert, L.~Roos, L.~Scotto~Lavina, A.~Tarek

\bigskip 
\bigskip
LPNHE, Sorbonne Universit{\'e}, Paris Diderot Sorbonne Paris Cit{\'e}, CNRS/IN2P3, Paris, France\\

\bigskip 
  \bigskip
  \end{center}
Contact:
\begin{itemize}
    \item[--] Vladimir Vava Gligorov: vladimir.gligorov@cern.ch
    \item[--] Gregorio Bernardi: gregorio@lpnhe.in2p3.fr
\end{itemize}

%\vspace{\fill}

\end{titlepage}

\pagestyle{empty}  % no page number for the title

%%%%%%%%%%%%%%%%%%%%%%%%%%%%%%%%
%%%%%  EOD OF TITLE PAGE  %%%%%%
%%%%%%%%%%%%%%%%%%%%%%%%%%%%%%%%
\pagestyle{plain}
\pagenumbering{arabic}
\setcounter{page}{1}

\onecolumn
%\twocolumn
\linewidth=\columnwidth

\section*{Reflections on the fundamental questions}

The past hundred years have seen giant leaps forward in our understanding of both the macroscopic and microscopic universe.
Nevertheless the highly predictive and largely accurate models developed in cosmology, astrophysics, and particle physics seem to be in
profound contradiction with one another. These contradictions are manifested for instance in the inexplicably large baryon asymmetry of the universe, the unexplained nature of dark matter (DM) and dark energy, and the unknown relationship between quantum mechanics and gravity.
Together they are a clear indication of particles and forces beyond our current understanding, and motivate the continued search for more complete theories of nature at both the very large and very small scales.  

The standard model of cosmology contains several ingredients which are required to explain current experimental observations. Ignoring the primordial singularity of the Big Bang and moving forward in time, they consist of inflation, which sets the initial conditions; DM, which explains the dynamics of the primordial baryon-photon fluid, the observed distribution of matter, and the dynamics of clusters and galaxies; and dark energy, which explains the luminosity distances to distant supernovae and the apparent size of the baryon acoustic oscillations feature observed in the distribution of matter.
The latest Planck measurements of the Hubble expansion rate and their apparent incompatibility with direct cosmic-ladder measurements also call into question the standard model of cosmology. Linking all of these topics is the question whether general relativity is an accurate description of gravity at all distance and energy scales. 

Among these questions, the nature of DM deserves special consideration, since it lies at the intersection of cosmology, astrophysics, and particle physics. Many theoretical models are compatible with current cosmological observations, and possible masses and couplings of the putative yet undetected particle ranges on tens of orders of magnitude. It is therefore essential to keep pursuing all possible methods for a DM detection: the interaction of DM particles with standard matter in dedicated ultra sensitive detectors placed on Earth; the production of DM in colliders or fixed-target experiments; and the observation of standard particles produced by the annihilation of DM in astroparticle experiments.

Understanding the sources of Ultra High Energy Cosmic Ray (UHECR) and mechanisms which accelerate particles up to $10^{20}$ eV opens another window on the interplay of macroscopic and microscopic phenomena. The discovery of gravitationnal waves (GW) has opened a new window onto this question, complementary to traditional messengers such as UHECR, photons and neutrinos. The effect of propagation on high energy particles in the cosmic backgrounds and the detailed studies of interactions of UHE particles with the atmosphere or other materials also test both our fundamental microscopic theories and our cosmological and astrophysical models.

Analogously to these astrophysical results, ground-based particle accelerators allow us to answer additional fundamental questions about the nature of neutrinos and the leptonic sector of the Standard Model (SM) of particle physics more generally. 
Oscillation experiments have allowed a first experimental grasp on neutrino masses which motivated a large program of accelerator and reactor based searches aiming at determining more precisely the (relative) neutrino mass pattern and the mass eigenstates mixing matrix parameters along with the CP violation aspects of this sector.
%The discovery of neutrino oscillations and therefore neutrino masses, using atmospheric and solar neutrinos, directly contradicted the SM hypothesis of massless neutrinos and has motivated a large programme of accelerator based measurements to understand the oscillation parameters, the neutrino mass hierarchy and the CP violation aspects of this sector. 
Cosmological observations can also bring insights about the neutrinos properties. Indeed the current best upper limits on neutrino masses come indirectly from the fraction of relic neutrinos and their impact on the small scale distribution of matter. Finally, the existence of CP violation in the neutrino sector would establish leptogenesis as a viable explanation for the observed asymmetry between matter and antimatter in the Universe.

The accelerator program has more generally the ambition to verify if the 17 particles and force-carriers of the SM are indeed elementary or rather composite; whether new particles or force-carriers exist with a mass and coupling within the range of ground-based accelerators; and if the properties of the Higgs boson, the only fundamental scalar of the model, agree with SM predictions. The level of agreement is both fundamental in establishing the nature of the electroweak symmetry breaking mechanism, and because Higgs boson properties are a highly sensitive probe into the mass scale of any physics beyond the SM. In particular a precise measurement of the Higgs self-coupling is key to understand the electroweak symmetry breaking mechanism. Flavour physics offers a complementary indirect window by addressing the validity of the SM picture of quark mixing (the CKM paradigm) and the validity of SM symmetries - for example baryon number or lepton flavour - in quark transitions. As with the measurements of Higgs properties, precision tests of these questions are sensitive to new particles or force carriers with a mass scale far beyond the ones available in ground-based accelerators.

%\subsection*{In Nuclear Physics}

\section*{Current scientific activities at LPNHE}

The LPNHE is engaged along three main axis, Cosmology, Astroparticles, and High energy Physics, that will be described in this order.

The cosmology group of LPNHE is part of several ongoing and upcoming experiments which address the fundamental questions : eBOSS, DESI and LSST. While eBOSS is scheduled to finish its data taking campaign in 2019, LSST and DESI will see first light starting in late-2019 (DESI) and 2020 (LSST). These experiments will obtain percent-level measurements of the history of the expansion rate through the matter-dominated to dark-energy dominated transition. They will deliver the strongest tests of dark energy models until the next generation of experiments, and LPNHE scientists will be at the forefront of the corresponding scientific analysis. Since 2015, the laboratory is also involved in searches for dark matter through direct detection, both using liquid noble gases (DarkSide and Xenon) and silicon (DAMIC). These mainly target WIMP models in an extended range of mass (from 1GeV to 1TeV), though others such as hidden sector based models can also be tested through more sophisticated analysis or an upgraded detector in the case of DAMIC.

LPNHE is also a historical contributor to the ground based gamma ray detector HESS and the Pierre Auger UHECR Observatory. The analysis of HESS data at LPNHE focused on the extragalactic source description and multi-messenger studies. Furthermore, the study of the time of flight of gamma rays from distant sources allowed to test a possible Lorentz invariance violation. LPNHE is also involved in DM searches in astrophysical objects or in the identification of anomalies in the diffuse spectrum. In recent years, these activities have been carried out in parallel with technical developments for the next generation of imaging Cherenkov telescopes the Cherenkov Telescope Array: CTA.
Beyond the early work on construction, the work on Auger has included contribution to source distribution studies (anisotropies) and more recently to the determination of the mass of UHECR thanks to radio detector developments and new analysis techniques. %The mass being a crucial parameter for the possible source pointing but also an important ingredient in the study of UHE particle interactions in the atmosphere.
%Since 2014, the laboratory is also involved in searches for DM through direct detection, both using liquid noble gases (DarkSide and Xenon) and silicon (DAMIC). These mainly target WIMP models in an extended range of mass (from 1GeV to 1TeV), though others such as hidden sector based models can also be tested through more sophisticated analysis or an upgraded detector in the case of DAMIC.

Within accelerator-based physics, the LPNHE neutrino group has been involved
for more than 10 years in the T2K long baseline neutrino oscillation experiment, with a precise measurement of $\theta_{13}$ and the first observation of $\nu_e$ appearance in a $\nu_\mu$ beam via oscillation. The next primary objective is to observe and characterize $CP$ violation in the neutrino sector.
%This experiment consists in sending an off-axis neutrino beam to a near detector called ND280 and a far detector called SuperKamiokande.
The $CP$ violating phase can be measured from the observation of electron neutrino and anti-neutrino appearance in a muon neutrino and anti-neutrino beam. The team contributes to the near detector upgrade for better charged particle tracking, to the oscillation analysis and to the hadron production experiment, NA61/SHINE, at CERN, which aims to precisely predict the T2K neutrino fluxes and reduce the systematic uncertainties on the CP violation measurement. 
In the sector of lepton flavour violation, the LPNHE is also involved in the COMET experiment which aims at observing neutrinoless conversions of muons to electrons, that are forbidden in the SM and would therefore be a clear signature of physics beyond the SM.

Within collider-based physics, LPNHE is today primarily involved in the ATLAS and LHCb experiments at CERN.
Both the current ATLAS and LHCb detectors have proven enormously successful, collecting over 160 and 9~fb$^{-1}$ of proton-proton collision data respectively. These data have enabled ATLAS to carry out a very broad physics programme, well beyond the discovery of the Higgs boson and the measurement of its properties. It encompasses precision measurements, in particular of the top and electroweak (EWK) SM sector and searches across the full range of physics scenarios (Supersymmetry, known also as SUSY, and non-SUSY) beyond the SM (BSM). The main activities of the ATLAS LPNHE group include the discovery and characterization of the Higgs boson in the di-photon and bottom quarks pair decays, precision tests of the SM in the cross-sections of jets, top quarks, and photons, and the search for dark matter candidates and new resonances. The group also contributes to improving the performance of the current detector and the 2024 upgrade (also known as Phase-II upgrade), with major responsibilities in the Inner Tracking (ITk), the High Granularity Timing Detector (HGTD) projects and the Hardware Track Trigger (HTT) for the ATLAS Trigger and Data Acquisition system upgrade.
LHCb has demonstrated the ability to make precise measurements of $CP$ asymmetries at twice the design instantaneous luminosity, as well as discover new exotic hadrons and study electroweak and heavy-ion phenomena in the forward direction. It has also made crucial measurements of rare decays as well as tree- and loop-level semileptonic decays of beauty hadrons, finding several deviations from the Standard Model and already with Run 1 data probing new particles with loop-level couplings at the scale of a few TeV, highly complementary to the direct searches from ATLAS. The LPNHE group has played a major role in several of the most important Run~1 analyses, in particular the angular analysis of $B\to K^{*0}\mu^+\mu^-$ and the test of muon-electron lepton universality in the same channel. The group is also contributing to Upgrade I of the LHCb experiment, due for the start of the 2021 datataking, with major responsibilities in the delivery of the tracker electronics, real-time reconstruction, and triggering.

It is also worth noting that LPNHE has made important contributions to the D0 and BaBar experiments in recent years, as well as the CALICE collaboration. Within D0, the group was deeply involved in the first evidence of fermion decays of the Higgs boson and contributed to measurements of the top quark properties. Within CALICE, the main contribution of the group consists of detector assembly with gluing and positioning robots, electrical tests of the sensors and metrology of detector units for a high granularity electromagnetic calorimeter Si-W. 

\section*{Perspectives for future scientific activities}

As LSST begins taking data, the LPNHE will exploit its expertise in CCD systematics, and supernovae analysis. The former is key and unavoidable to properly interpret weak lensing measurements. The latter will lead to a novel velocity survey based on residuals to the Hubble diagram. The road-map of the cosmology group in the 5-year scale is clear and exciting: strong presence in commissioning activities followed by leadership of key scientific analyses. The challenges that the group will face will be  related to the question of how to optimally exploit the synergies and complementarities of its members and the experiments they participate to. In this context, the group is currently considering participating to the Prime Focus Spectrograph, a follow-up of Subaru/HSC scheduled to begin operations in 2021, and 4MOST, a spectrograph survey that will follow-up on LSST. 

In the direct continuation of HESS, the gamma ray group will mainly focus on CTA. CTA is planned to be commissioned in the year 2020 and will provide observations for 30 years. The technical activities in CTA will continue until the official delivery and the scientific involvement is foreseen for at least a decade. 
For UHECR and UHE neutrinos the LPNHE is involved in the Auger upgrade Auger Prime and the radio extension Auger Horizon. Interest is also expressed for the UHE neutrino and UHECR radio experiment GRAND and its first step GRANDProto300 expected to be deployed in 2020. GRAND would contribute to the multi-messenger effort and have a large impact on its own if technical challenges can be overcome.
% This will be tested with the GRANDProto300 experiment, which is expected to be deployed in 2020.
Finally, the space based GW experiment LISA is of prime interest for the multi-messenger studies since GW detection could happen up to one month before the counterpart in other messengers. Therefore, the question of an implication in this field is open.

For direct DM searches, liquid noble gas experiments such as Xenon and DarkSide to which the LPNHE is contributing in will eventually improve their sensitivity down to the point where neutrino coherent scattering with nuclei become dominant (the so-called ``neutrino floor'').
Here the Dark Matter group is also involved in two future large scale experiments: the 50 ton Xenon detector DARWIN and the 300 ton liquid Argon detector Argo.
% in the 300 ton LAr detector Argo and in the future very large Xenon detector, DARWIN. 
The upgrade of DAMIC, DAMIC-M is being developed and has a defined timeline for the next 5 years.
The ultra sensitive DM detector can be used for other physics analyses, for instance in the case of noble gas detectors for solar neutrino studies or searches of neutrino-less double beta decay.
Other well-motivated models based on a hidden sector or ultra light dark matter can be probed with small scale experiments implementing various techniques such as the use of graphene or nuclear magnetic resonance to search for axions (CASPEr).
Involvement in these types of experiments would allow LPNHE to make a large contribution relative to the overall experiment size, and possibly to collaborate with other labs of the Sorbonne Universit\'e campus. 

%If no DM signal were found, two choices would arise:
%\begin{itemize}
    %\item [-] The ultra sensitive DM detector can be used for other physics analyses, for instance in the case of noble-gas detectors for solar neutrino studies or searches of neutrino-less double beta decay.
    %\item[-] Other techniques are developed and employed to probe other models. Specific ranges of mass in DM models based on a hidden sector or ultra light dark matter (including axion) can be achieved with table-top experiments. Involvement in these types of experiments would allow LPNHE to make a large contribution relative to the overall experiment size, and possibly to collaborate with other labs of the Sorbonne Universit\'e campus.
%\end{itemize}

In line with its past and current activities, the Neutrino group will participate in HyperKamiokande. This next generation long baseline neutrino oscillation experiment aims at excluding CP conservation (a.k.a. $\delta _{\mathrm{CP}}=0$ or $\pi$) with a $5~\sigma$ significance. To achieve such sensitivity, a water Cherenkov tank about ten times larger than the existing SuperKamiokande far detector will be equipped with photomultipliers.
The construction of this detector will start in 2020 with a data taking start expected in 2026.
Beyond the measurements of neutrino oscillations, HyperKamiokande will have unprecedented sensitivity to the proton decay, by extending previous limits by at least one order of magnitude to $Br /\tau <10^{35}~\mathrm{years}$, and will be the most sensitive neutrino observatory for several astrophysical phenomena such as solar neutrinos, supernova explosions or supernova remnants.
Finally in the context of developing the next generation neutrino detectors for these two experiments, the group continues its contributions to the CERN Neutrino Platform for the T2K near detector upgrade and potentially the testing of photomultipliers for HyperKamiokande.

In terms of collider physics, the ATLAS and LHCb groups have well defined activities and plans through Run~3 of the LHC (ending 2023) and into the HL-LHC period which will begin around 2026. The LPHNE-ATLAS team is commited to two already approved Phase-II upgrade projects: ITk and HGTD. The group expertise in tracking detectors and jets and photons performance will be essential to exploit the physics potential of HL-LHC. The LHCb group's first priority is to successfully deliver the LHCb Upgrade~I,
which will take data during Run~3 and the first part of the HL-LHC period (Run~4).
The group is also motivated to participate in the R\&D for the LHCb Upgrade~II,
and members of this group played a major part in writing and editing the physics case for such
an upgrade. Achieving the desired luminosity of $1-2\cdot 10^{34}$cm$^{-2}$s$^{-1}$, or $5-10$ times
that of the LHCb Upgrade~I, will require major advances in both the tracking detectors, DAQ, triggering,
and reconstruction, topics which are well matched to the group's expertise. Because LHCb is a general purpose detector,
the technical expertise gained during this R\&D program will also be applicable to other experiments and
facilities, whether or not Upgrade~2 is approved.

\begin{table}
%\begin{small}
\caption{\label{tab:machines} Summary of tentative physics performance and timelines for collider experiments future projects.}
\begin{center}
\begin{tabular}{ l c c c c r }

Collider&	 type&	$\sqrt{s}$ &	Luminosity &	Starting date&	End date \\
\hline
\hline
HL-LHC &	pp circular&	14 TeV&	6-8 ab$^{-1}$&	2026&	2036\\
\hline
HE-LHC&	pp circular&	27 TeV&	30 ab$^{-1}$&	2040&	2050\\
\hline
LHeC&	ep \ hybrid&	1.3 TeV&	~1 ab$^{-1}$&	2032&	2045\\
\hline
\multirow{2}{*}{ILC} & \multirow{2}{*}{ee \ linear} & 0.25 TeV & 2 ab$^{-1}$ & {2034} & 2049 \\
&   & 0.50 TeV & 4 ab$^{-1}$ & 2050 & 2060 \\
\hline
\multirow{3}{*}{CEPC} & \multirow{3}{*}{ee circular} & 0.09 TeV & 8 ab$^{-1}$ & {2037} & 2039 \\
%\multirow{3}{*}{2040 (2yr@ Z; 1@WW; 7@ZH)} \\
&   & 0.16 TeV & 2.6 ab$^{-1}$  & 2039& 2040 \\
&   & 0.24 TeV & 5.6 ab$^{-1}$ & 2030 & 2036 \\
\hline
\multirow{4}{*}{FCC-ee} & \multirow{4}{*}{ee circular} & 0.09 TeV & 150 ab$^{-1}$ &   2039 &  2042 \\
&   & 0.16 TeV & 10  ab$^{-1}$           & 2043 & 2044\\
&   & 0.24 TeV & 5.0 ab$^{-1}$   & 2045 & 2047\\
&   & 0.36 TeV & 1.7 ab$^{-1}$ & 2049 & 2053\\
\hline
FCC-hh &	pp circular&	100 TeV&	25 ab$^{-1}$&	2043 if no FCC-ee&	2065\\
\hline
FCC-eh&	ep \  hybrid&	3.7 TeV&	~8 ab$^{-1}$&        2043 if no FCC-ee& 2065\\
\hline
SPPC&	pp circular&	75 TeV&	~5 ab$^{-1}$&	~2045&	~2060\\
\hline
\multirow{3}{*}{CLIC} & \multirow{3}{*}{ee \ linear} & 0.5 TeV & 0.2 ab$^{-1}$ & \multirow{3}{*}{2035} & \multirow{3}{*}{2050} \\
&   & 1.5 TeV & 0.4 ab$^{-1}$  & & \\
&   & 3.0 TeV & 0.7 ab$^{-1}$   & & \\
\hline
\hline

\end{tabular}
\end{center}
%\end{small}
\end{table}

Beyond this point LPNHE activities on collider experiments are closely tied to the availability and timelines of planned new experiments and facilities. In order to help put our thoughts about these in context, Table~\ref{tab:machines} compares their tentative physics performance and timelines, with no attempt to compare costs or otherwise comment on feasibility. Given these timescales, we can make some general observations about the scientific interest in the different facilities, dividing the experiments broadly into hadronic and leptonic colliders. We note upfront that an important advantage of $e^+e^-$ machines is the low pile-up and underlying event, so that the predictions are much less dependent on higher order QCD radiative corrections than in $pp$ colliders. When considering the future facilities we also stress the continued long-term importance of both direct searches for BSM particles and indirect searches for BSM effects in e.g. flavour physics, Higgs couplings, or other precision tests of the SM.

HE-LHC is the most natural extension of the HL-LHC program, since it would mostly involve replacing magnets in the current LHC tunnel.
The main interest of HE-LHC for Higgs boson physics is to explore a new energy frontier and measure the Higgs self-coupling to 20\% on the timescale of ~2050; other Higgs boson couplings might be systematically limited after HL-LHC. The increase in the center of mass energy opens some possibilities for new particle discoveries. This might be particularly interesting if the current flavour physics anomalies are confirmed, since most theoretical explanations place the new physics at the scale of a few TeV. Regarding the study of DM produced at colliders, there are systematic limitations for many scenarios.

The ILC baseline option ($\sqrt{s}=250$~GeV) would allow precise Higgs, top and EWK coupling measurements by 2050, but no Higgs self-coupling measurement. For this the machine would have to be extended to 500~GeV, which would allow to reach 27\% uncertainty by around 2060. The direct discovery potential for new physics is limited, however if LHC discovers physics beyond the SM then ILC could be a precision microscope for characterizing it.  
 
With another linear approach, CLIC delivers less precise Higgs couplings than ILC on the same or longer timescales, while the potential prospects for direct discovery of new physics depend on the scenario considered. 
 
 In contrast with the ILC or CLIC, the main motivation for FCC-ee is the precise study of the complete electroweak sector, starting with huge statistics of Z decays, then WW production, followed by an extensive Higgs program which will lead to the most precise Higgs couplings measurement by a similar timeline (2047 or so). These would be even significantly improved by the foreseen additional 5 years at 350-365 GeV, which would also lead to a Higgs boson self-coupling measurement with 40\% uncertainty by 2053. It would also enable a unique programme of top quark physics, in particular the measurement of the pole top mass, which is fundamentally connected to the vacuum stability of the universe. The Z pole programme would also guarantee access to an unprecedented amount of b-hadrons hence an exceptional potential for flavour physics and other indirect BSM probes. After FCC-ee, the tunnel could be reused for FCC-hh, at 100 TeV or even more if developments of magnets would have progressed further, with the potential
to measure the Higgs self-coupling to 5\% and largely extend the reach for BSM searches, including DM collider production.

The considerations for CEPC/SPPC are quite similar to FCC-ee/hh. The CEPC schedule is more aggressive (potential start in 2030) but it will run at lower instantaneous luminosity, and in the baseline design the $e^+e^-$ center of mass energy is limited to 240~GeV, hence it will not allow to reach sub-percent uncertainty on the couplings. The CEPC schedule also tends to collide with the HL-LHC effort. %Reina: discuss with other members of the team. 

% The potential for direct discovery of new physics depend on the scenario considered. %For example a precision of O(1\%) is found on the mass measurement of different sparticles across different channels and SUSY models.  

Finally, LHeC has the potential of delivering complementary measurements of light quark and EW couplings, and allowing to reach a higher center of mass energy than an $e^+e^-$ machine (1.3 TeV for LHeC, 3.7 TeV for pp-eh). It is often claimed that LHeC can allow an $\alpha_S(m_Z)$ measurement with 0.1\% level precision. Compared to an $e^+e^-$ machine the lepton-hadron environment offers similarly clean signals (no pile-up) but with lower luminosities. Regarding BSM scenarios it is important to notice the sensitivity of these machines to eeqq contact interactions and electron-quark resonances. The electron beam is provided by a 3-pass recirculating energy-recovery linac (ERL) which can be built without strong interference with the proton machine, and which can be used for both LHeC and FCC-ep configurations, meaning that LHeC/FCC-ep would complement the LHC/FCC CERN program at reasonable cost without disrupting it.

It is also important to mention that out of mainstream and/or smaller experiments focused on understanding a particular sector of the SM or performing specific searches for new physics could also be interesting for the next few years. For example, the recent demonstration of muon ionisation-cooling by the MICE Collaboration opens a path to a neutrino factory or a muon collider. Regarding smaller experiments we can give as examples, the SHip (Search for Hidden Particles) experiment, a beam dump facility to be built at CERN and to search for weakly coupled particles in the few GeV mass range, CODEX-b which is a proposed extension of the LHCb experiment to search for long-lived particles in the old DELPHI cavern, or MATHUSLA (Massive Timing Hodoscope for Ultra Stable Neutral Particles), a surface detector placed 100 metres above either ATLAS or CMS to detect heavy long-lived particles.

%Muons are roughly 200 times heavier than the electron and thus emits around 10$^9$ times than an electron beam of the same energy, it is also possible to produce multi-TeV collisions in an LHC-sized circular collider. The large muon mass also enhances the $S$-channel Higgs production rate by a factor of around 40000 compared to that in $e^-e^+$  colliders, making it possible to scan the centre-of-mass energy to measure the Higgs-boson line shape directly and to search for closely spaced states

\section*{Conclusions and recommendations}

The short-medium term future of high energy physics at colliders is well established with the HL-LHC program which is expected to bring a wealth of important results on the Higgs boson, on the standard model more generally, on the exploitation of flavour physics and hopefully on physics beyond it through direct or indirect searches. We would like to stress that the full exploitation of the HL-LHC physics programme requires the second upgrade of LHCb to be built.

For future machines, we are waiting for a positive answer of the Japanese government for launching ILC. The LPNHE, which has been involved in it since many years would continue and increase its involvement if a positive answer comes in a timely way. But for us, the main question for the future of the field is if CERN goes towards HE-LHC, or if it goes directly towards building a circular 100 km tunnel, allowing for FCC-ee and later for FCC-pp. 

We note that the HE-LHC program would delay significantly the most precise measurement of the Higgs boson couplings reachable at FCC, in which signs of BSM physics could be seen, while the Higgs boson self-coupling measurement can also be obtained in the FCC-pp option later but with better precision. The direct searches for physics beyond the standard model would also be done significantly better in the long term at FCC-pp. The CEPC/SPPC option would provide a similar program as the FCC-ee/pp, but with reduced goals and sensitivities, and a shorter timescale overlapping with the activities on the HL-LHC. Its construction would nevertheless be a most welcome step towards our common goals.

The most precise study of the Higgs boson which the FCC program allows (couplings and decays to invisible final states with the ee machine, then self-coupling and rare decays with the pp machine), and the highest energy frontier which will eventually be reached, appears to us the most promising route that colliders can take towards physics beyond the standard model (BSM). FCC-ee also has an outstanding W, top, and Z pole program, the latter also presenting an exceptional potential for flavour physics and other indirect BSM probes. 

Performing specific searches should also be encouraged and smaller experiments, such as SHiP, CODEX-b, or MATHUSLA would complement nicely a flagship program.

On the neutrino side, the LPNHE will continue its involvement in the long baseline oscillation experiments based in Japan, from T2K-II to HyperKamiokande. 
The related detector developments will be conducted in the CERN Neutrino Platform.

In conclusion, given the current HL-LHC program, reach and schedule, the FCC program (ee then pp) appears to us as the most promising future path to reach deeper understanding of elementary particle physics, while other collider and non-collider future facilities would also be very rewarding. Overall, our involvements in the next generation experiments in the fields of Particle Physics, Astroparticles and Cosmology, will allow our Laboratory to participate in what are likely to be profound and exciting discoveries.

\end{document}